\documentclass{llncs}
\usepackage{epsf}
\usepackage{booktabs}
\usepackage{pgfplots}
\usepackage{amsmath}
\usepackage{mathtools}
\usepackage[hyphens,spaces,obeyspaces]{url}
\usepackage{array}
\usepackage[hidelinks]{hyperref}
\usepackage{comment}
\usepackage[space]{cite}
\usepackage{url}

\urldef{\mailsa}\path|{ahsaas.bajaj, shubham.k1, mukund.r, h.tiwari, vanraj.vala}@samsung.com|

\newcommand{\repeatthanks}{\textsuperscript{\thefootnote}}
\begin{document}
\pagestyle{empty}
\title{RelEmb: A relevance-based \\application embedding for Mobile App\\ retrieval and categorization}
\author{Ahsaas Bajaj\thanks{indicates equal contribution}, Shubham Krishna\repeatthanks, Mukund Rungta, \\Hemant Tiwari and Vanraj Vala}
\institute{Samsung R\&D Institute, Bengaluru, India \\ \mailsa}
\maketitle
\begin{abstract}
Information Retrieval Systems have revolutionized the organization and extraction of Information. In recent years, mobile applications (apps) have become primary tools of collecting and disseminating information. However, limited research is available on how to retrieve and organize mobile apps on users' devices. In this paper, authors propose a novel method to estimate app-embeddings which are then applied to tasks like app clustering, classification, and retrieval. Usage of app-embedding for query expansion, nearest neighbor analysis enables unique and interesting use cases to enhance end-user experience with mobile apps.
\end{abstract}
\section{Introduction}
Recent years have seen tremendous increase in usage of mobile applications, a.k.a. \textit{apps}, mainly due to the ever rising popularity of smart phones. There are millions of apps \cite{statista} freely available on Google Play Store and Apple App Store. With the abundance of data available in form of applications, it is important to build efficient and effective retrieval engines around them. When a user queries for an application, it is crucial to bring the most relevant application at the top. In a mobile environment, the user expects search operation to be rapid and relevant. This is a challenge as the number of available applications are readily growing these days. Most of the recent work in the field of Information Retrieval has been focused on web-search scenarios and improving the ranking of web results. Little work is focused on information retrieval centered around mobile applications. Intelligent retrieval methods are required to make sense of this large amount of app data and also keep them organized in users' devices. Since most of the data related to applications are in the form of its description, it is important to mine this source of information. Very recently, neural word embedding has found its use in the field of Information Retrieval. A novel method of learning word embedding from app description is proposed in this paper. The paper is organized as follows. 

An overview of related work in the domain of information retrieval is presented in Section~\ref{relatedWork}. The details of estimating the embeddings are discussed in section~\ref{proposed}. Section~\ref{experiment} elaborates experiments and results using the proposed embeddings for various tasks. Finally section~\ref{conclusion} recapitulates the proposed approach with current applications and scope of future extensions.
 
\section{Related Work}
\label{relatedWork}
In past few years, word embeddings have found a major role for solving different tasks in multiple domains like Natural Language Processing and Machine Learning. Word embedding is basically a representation of a word in a vector space and there are different ways to learn this representation- to model semantic or syntagmatic relationships. Traditionally, techniques like Singular Value Decomposition (SVD) and Principal Component Analysis (PCA) were applied to generate dense word representations. With the advent of neural networks, it became possible to learn more enhanced word representations. Different variants of word embeddings such as word2vec \cite{mikolov2013distributed}, GloVe \cite{pennington2014glove} have been proposed for learning dense distributed representation of words. These window based methods count the number of co-occurrences of terms in the training corpus and suggest how likely is a term to occur with other terms. Such methods have been highly effective in solving different problems such as Text Summarization, Sentiment Analysis, and Machine Translation. Later, the idea of neural word embeddings was extended to learn document-level embeddings \cite{le2014distributed} which opened further interesting use-cases. Information Retrieval is one such domain.

In 2015, a unified framework for monolingual and cross lingual information retrieval was formulated using word embeddings \cite {vulic2015monolingual}. Dual Embedding Space model using word embeddings was proposed for the task of document ranking \cite{mitra2016dual}. Also, there are methods for solving the problem of term weighting and query interpretation using distributed representation of words \cite{zheng2015learning}. Word and document embeddings have also found their use in applications such as query expansion and estimation of accurate query language models in the task of retrieval \cite{zamani2016embedding}. In general, query expansion is based on pseudo relevance feedback \cite{rocchio1971relevance} to enhance the performance of ad-hoc retrieval. Initial work in query expansion utilized word2vec (continuous bag-of-words) embedding approach \cite{kuzi2016query}. Currently, researchers are experimenting to build robust word embeddings to analyze their usage in different problems of information retrieval. The novelty of  locally-trained word embeddings are nicely highlighted in \cite{diaz2016query}. They have shown that these locally trained word embeddings outperform the globally trained ones (like word2vec, GloVe, etc.) for retrieval tasks. It isn't necessary that what a word means globally also means the same in a local context. Building on the same line of work, relevance-based word embeddings \cite{zamani2017relevance} were explored to propose two new cost functions to learn word embedding specifically for retrieval tasks like query expansion and classification. 

These days, mobile applications form a large chunk of data which is available for consumption. Some work has been done to mine user-reviews on mobile applications \cite{vu2015mining}. Mobile app retrieval has been experimented by few researchers to learn application representation using topic modeling \cite{park2015leveraging} and intention in social media text \cite{park2016mobile}. Very recent work has been done to build a unified mobile search framework using neural approaches to learn high-dimensional representations for mobile apps \cite{Aliannejadi:2018:TAS:3209978.3210039}. Also, there are recent attempts made to use Learning-to-Rank algorithms for optimizing retrieval of mobile notifications triggered by installed apps \cite{bajaj2019enhanced}. However, application embeddings have been rarely utilized to perform query expansion, nearest neighbor analysis and other tasks related to mobile applications. 

\begin{sloppypar}This paper proposes a novel method to learn dense word embeddings from app descriptions. The learned word embeddings are then used to compute relevance-based application embeddings which are suitable for retrieval and categorization of Mobile Apps. 
\end{sloppypar}
\section{Proposed Method}\label{proposed}
The following section details the method to extract word embeddings from app descriptions. As discussed in section~\ref{relatedWork}, embedding techniques like word2vec and Glove are not suitable for information retrieval tasks as they are based on co-occurrence and fail to capture the associated relevance. It is observed that task specific embeddings generally perform better. To improve performance in tasks related to retrieval, it is important to learn these representations which carries the notion of relevance. With this motivation, the following subsection~\ref{wordemb} discusses the process of learning this representation (word embedding) for each unique word in the vocabulary. Once this sparse representation is calculated, it is used for learning a distributed vector representation using neural networks. This reduces noise, identifies patterns and is suitable for mobile devices as computations involving dense vectors are faster and requires less space for storage. The approach proposed in this paper is language-agnostic as while formulating the method it is assumed that application descriptions can be of any language and is seen as collection of words. Therefore, same techniques can be applied if datasets for different languages are available for training the models. For the methods discussed below, the terms - application embedding and document embedding are used in the same context. 
\subsection{Relevance-Based Word Representation}
\label{wordemb}
As it is said in linguistic theory \cite{firth1957synopsis}: ``You shall know a word by the company it keeps". For the domain of information retrieval, it can be said that a word is known by the documents (or other words) it retrieves. In order to capture the relevance associated with each word, the information contained in the top retrieved documents are utilized when the same word is used as a query to retrieval engine. Apache Lucene \cite{Lucene}, an open source search library is used to test this hypothesis and get the relevant results for each word in the vocabulary. Lucene uses Okapi-BM25 score \cite{robertson1995okapi} to rank the retrieved documents. Using the app descriptions for all applications, Vector Space Model (VSM) representation is constructed using the term frequency \cite{luhn1957statistical} and inverse document frequency \cite{sparck1972statistical} (tf-idf) scores. 

In this manner, every application is represented by a vector of dimension $1 \times N$, where $N$ is the vocabulary size of the corpus. The weight vector for application $j$ is defined as \cite{salton1975vector}:
\begin{equation}
\vec{VSM_{j}} = [w_{1,j}, w_{2,j},..., w_{N,j}]
\end{equation}
\begin{equation}
w_{ij} = tf(i,j) \cdot \log \frac{Total Docs}{\mid \{ j' \epsilon Doc \mid i \epsilon j' \} \mid}
\end{equation}
$$ tf(i,j) \text{ is the term frequency of term } i \text{ in application } j $$
$$ TotalDocs \text{ is the total number of documents (applications) in the corpus} $$
$$ \mid \{ j' \epsilon Doc \mid i \epsilon j' \} \mid \text{ is the total number of applications containing the term } i $$
$$ \log \frac{Total Docs}{\mid \{ j' \epsilon Doc \mid i \epsilon j' \} \mid} \text{ is inverse document frequency.}$$
\newline
For a query $t$ with words $\{t_1, t_2,..,t_{n}\}$, BM25 score of document (application $j$) is computed as:
\begin{equation}
BM25(t,j) = \sum_{i=1}^{n} idf(t_i) \cdot \frac{tf(t_i,j)\cdot (k_1 + 1)}{tf(t_i,j) + k_1 \cdot (1-b+b\cdot\frac{\mid j \mid}{avgdl})}
\end{equation}
$$ \mid j \mid \text{ is the length of document $Doc$ in words} $$
$$ avgdl \text{ is the average document length} $$
$$ k_1 \text{ is the hyper-parameter that calibrates the scaling of } tf(t_i,j)$$
$$ b \text{ is another tuning parameter which determines the scaling for } avgdl$$

\subsubsection{Method} Firstly, the most relevant documents ($topDocs$) are retrieved after querying the system with each term in the vocabulary. The number of top documents is an input provided to the system and is given by the variable $len(topDocs)$ . For the experiments in this paper $len(topDocs)$ is set to 10. Using the constructed $VSM$ and $BM25$ scores for these $topDocs$, a high-dimensional word representation is computed by the following equation:
\begin{equation}
\label{wordrepr}
\vec{WordRepr_k} = \frac{\sum_{i=0}^{len(topDocs)} BM25_{i,k} \cdot \vec{VSM_i}}{\sum_{i=0}^{len(topDocs)}BM25_{i,k}}
\end{equation}
$VSM_i$ is the Vector Space Model representation for the $i^{th}$  application and $BM25_{i,k}$ is the BM25 score of the $i^{th}$ document for term $k$.
In this manner, word embeddings of all words in the vocabulary are computed. This word embedding takes into account the relevance of word in accordance with the context of data.
\subsection{Dense Word Embedding}
\label{relemb}
Curse of Dimensionality is a well known phenomenon in the field of Machine Learning and Data Mining. After a certain point, the increase in number of dimensions hurts the performance of algorithms. Inferring and performing computations on sparse vectors and matrices are costly operations as they may contain noise and irrelevant information. Even the vector given by equation~\ref{wordrepr} is sparse in nature which makes its usage limited. Hence, dimensionality reduction is used to remove noise and extract useful patterns. Different algorithms like Principal Component Analysis, Singular Value Decomposition, and Neural Networks are being used for performing the task of dimensionality reduction.

\begin{figure}[h]
	\centering
	\includegraphics[width = 10cm, height = 8cm]{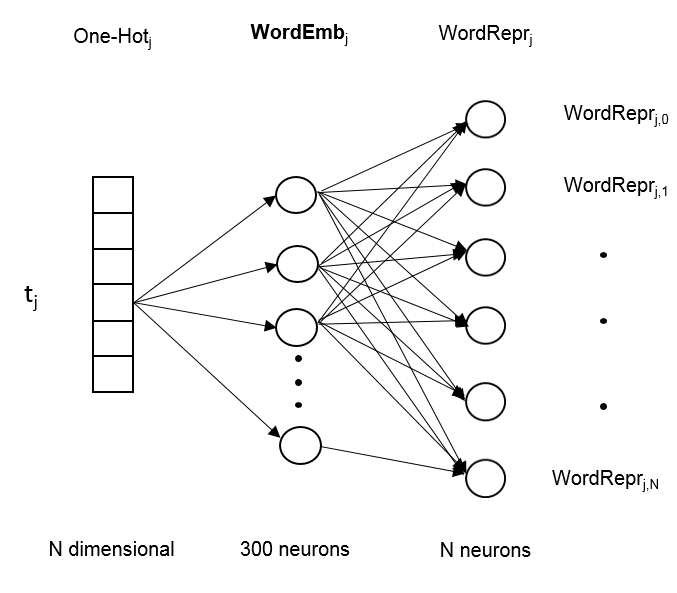}
	\caption{Neural architecture to generate Word Embedding from Word Representation}
	\label{fig:neural}
\end{figure}

Firstly, to learn word embeddings from word representations, a simple yet effective feed-forward neural network architecture is used. The architecture is shown in Figure~\ref{fig:neural}. The input fed to the network is a N dimensional one hot encoded vector, where N is the vocabulary size and the output layer is N dimensional word representation calculated in equation~\ref{wordrepr}. The activation functions used in the hidden and the output layer are ReLU and softmax respectively. Adam Optimizer has been used for updating the parameters of the neural network and the loss function used is cosine distance. The neural network tries to learn latent patterns present in the word representation while reducing dimension, making it computationally faster and cheaper for performing different retrieval tasks. For training the word embeddings, unique words from the app descriptions are used and passed as one hot encoded vector (input vector). The corresponding word representation is fed as the output of neural architecture. The dimension of the hidden layer is set to 300. Tensorflow framework \cite{Tensorflow} is used for the training process. For vocabulary size $N$, let the two weight matrices be $W_1$ and $W_2$ and $t_j$ be the one-hot encoding of term $j$ (input vector). 
\begin{equation}
\label{wordEmbOneHot}
\vec{WordEmb_j} = ReLu(\vec{t_j} \times W_1)
\end{equation}

where $W_1$ has the dimension of $N \times 300$ and $ReLu(x) = max(0,x)$ as given in \cite{glorot2011deep}. The output layer is given by:
\begin{equation}
\label{softmax}
softmax(\vec{WordEmb_j} \times W_2 + \vec{bias}) 
\end{equation}
$W_2$ has the dimension of $300 \times N$ and $softmax$ is the activation function \cite{bridle1990probabilistic}. $bias$ is a vector of dimension $1 \times N$. The cosine distance loss is computed against the values from equation~\ref{softmax} and the representation stored in $WordRepr_j$. After training, the hidden layer gives the word embedding $WordEmb$ for term $j$.
This relevance based embedding has been developed keeping in mind the notion of relevance. This embedding is not intuitively built for tasks such as classification and clustering. To tackle such problems, a vanilla auto-encoder is trained to learn the word embeddings from the word representations proposed in equation~\ref{wordrepr}. The input and output to this auto-encoder are the word representations given by equation~\ref{wordrepr}. The embeddings learned in the hidden layer of the auto-encoder are denoted by $WordEmbAE$ for the rest of this paper. 

With the discussed $WordEmb_j$ representation for each term $j$ in vocabulary, an embedding matrix is constructed.
\subsection{RelEmb: Application Embedding}
\label{docemb}
The intention behind any user query is to get the most relevant applications. Application title or category is not sufficient to extract relevant applications. But, each application has an associated description, which carries the useful information about the application and its features. This description can be considered as bag of words and can be used to generate word embeddings as discussed in sections~\ref{wordemb} and~\ref{relemb}. The learned word embeddings can be extended for calculating application-level embedding by aggregating the word embedding of each term in the application's description.  Let an application $k$ (document) be represented as $description(k)$ = \{ $w_1, w_2, w_3,...,w_i,..., w_n \}$ where each $w_i$ is the word from vocabulary.
\begin{equation}
\label{eq:relemb}
RelEmb_{k} = \frac{\sum_{i=0}^{n} WordEmb[w_i]}{n}
\end{equation}
Similarly, when the word embeddings are learned using vanilla auto-encoder, application embeddings are given by:
\begin{equation}
RelEmbAE_{k} = \frac{\sum_{i=0}^{n} WordEmbAE[w_i]}{n}
\end{equation}
\section{Experimental Details}
\label{experiment}
To evaluate the performance of the proposed method, extensive experiments are performed for different tasks related to application retrieval and categorization. A publicly available apps dataset is used for testing the performance of the embeddings. The dataset includes the data for query-application relevance judgment, which is useful to test the retrieval task of Query Expansion. A qualitative experiment in terms of nearest neighbor analysis is also elaborated in this paper. This shows the capability of app embedding for tasks like application recommendation. The dataset also includes a category tag for each application which can be used to analyze the performance of application embeddings for the task of multi-class app classification. Apart from supervised learning, these embedding vectors can also aid in unsupervised tasks like app clustering. The  evaluation of embeddings has been discussed in this section. Results indicate superior performance as compared to existing state-of-the-art methods for various tasks.
\subsection{Dataset}
\label{dataset}

Data Set for Mobile App Retrieval \cite{park2015leveraging} (TIMAN dataset) is used for evaluation of the proposed methods. This data includes information about 43,041 mobile apps including the title, description, category, package name, user-reviews, query-document relevance pairs, etc. To trim down the vocabulary size, only English words from the app descriptions are selected. Moreover, minimum permissible length for each word was set to two. With the above mentioned preprocessing, the number of unique apps comes down to 42,895 with a vocabulary size of 37,213.
\subsection{Application Retrieval}
\label{retrieval}
In mobile apps scenario, user generally queries with short-text, mostly containing 1-2 terms. It becomes a challenge for any retrieval engine to bring the most relevant results at the top ranks. Solution like query expansion helps to re-formulate the user query which eventually improves the retrieval accuracy. But, it is also problematic to bring the useful expanded terms which increases the relevance of the results. The word embeddings which are discussed in sections~\ref{wordemb} and~\ref{relemb} are based on pseudo-relevance feedback (BM25) and are trained in a manner suitable for retrieval tasks. Authors believe that these embeddings are capable of finding the best expansion terms for user-query. 
\subsubsection{Evaluation with Query Expansion}
\label{QE}
Query expansion is a standard retrieval task which has its practical advantages. It boosts the performance of a retrieval engine using a two-pass method. Query expansion approach is also an appropriate methodology to validate the performance of a retrieval system. Using the word embeddings proposed in section~\ref{proposed}, query expansions tasks are evaluated. TIMAN dataset also provides the query-application relevance data with 4533 instances. The relevance score is a real-valued number in the range of 0-2. Since, it is a floating point (and not just binary) relevance label, $NDCG$ metric \cite{jarvelin2002cumulated} is used to judge the effectiveness of query expansion. The number of expanded terms is given by the variable $k$. To find the expansion terms for a given query, the following methodology is used.
\begin{itemize}
	\item Calculation of input query vector using multiple terms in the user-query $query = \{ w_1, w_2, ...\}$
	$$ \vec{Q} = \frac{\sum_{i=0}^{len(query)} WordEmb[w_i]}{len(query)} $$
	where, $WordEmb[w_i]$ are calculated in equation~\ref{wordEmbOneHot}.
	
	\item Given the query vector $\vec{Q}$, the most semantically similar terms are found by calculating cosine similarity on the $WordEmb$ matrix.
	
	\item Top $k$ terms from the $WordEmb$ matrix are selected based on their values of cosine similarity with respect to the input query vector $\vec{Q}$.
	
	\item Example: $query = \{music, stream, airplay\}$ retrieves top five expansion terms as $\{sonos, bose, dlna, player, listen\}$. It can be seen that these terms are closely related to the initial user query and all of them deal with music and streaming services.
\end{itemize}
\begin{table*}
	\caption{Evaluation results on Query Expansion}
	\label{table:QueryExpansion}
	\centering	
	\begin{tabular}{>{\centering\arraybackslash}p{18mm}>{\centering\arraybackslash}p{18mm}>{\centering\arraybackslash}p{18mm}>{\centering\arraybackslash}p{18mm}>{\centering\arraybackslash}p{18mm}>{\centering\arraybackslash}p{18mm}}
		\toprule
		& \multicolumn{5}{c}{\textbf{Expansion Techniques (k=5)}}\\
		\cmidrule{3-6}
		\textbf{Metric} & Okapi-BM25 & SVD & PCA & WordEmbAE & WordEmb\\
		\midrule
		NDCG@3     &0.569 &0.499 &0.467 &0.572 & \textbf{0.577}  \\
		NDCG@5     &0.542 &0.502 &0.469 &0.548 & \textbf{0.563}  \\
		NDCG@10    &0.535 &0.505 &0.477 &0.542 & \textbf{0.556} \\
		\bottomrule
	\end{tabular}
\end{table*}
Query expansion results are shown in Table~\ref{table:QueryExpansion}. Okapi-BM25 scores are used as baseline by considering the input query $\vec{Q}$ as the direct query to the retrieval system (without expansion). The novelty of the proposed word embeddings can be seen from the increase in NDCG scores after query expansion.

$WordEmb$ (equation~\ref{wordEmbOneHot}) performed better than $WordEmbAE$ as the intuition behind learning these embedding was the notion of relevance for the tasks in information retrieval. Query expansion is the right measure to validate this claim and the results fully support it. Even then, $WordEmbAE$ was able to out-perform the baseline BM25 scores, proving the benefits of the sparse representations discussed in section~\ref{wordemb}.
\subsubsection{Nearest Neighbor Analysis}
\label{NNA}
\begin{figure}[h]
	\centering
	\includegraphics[width=12cm,height=6cm]{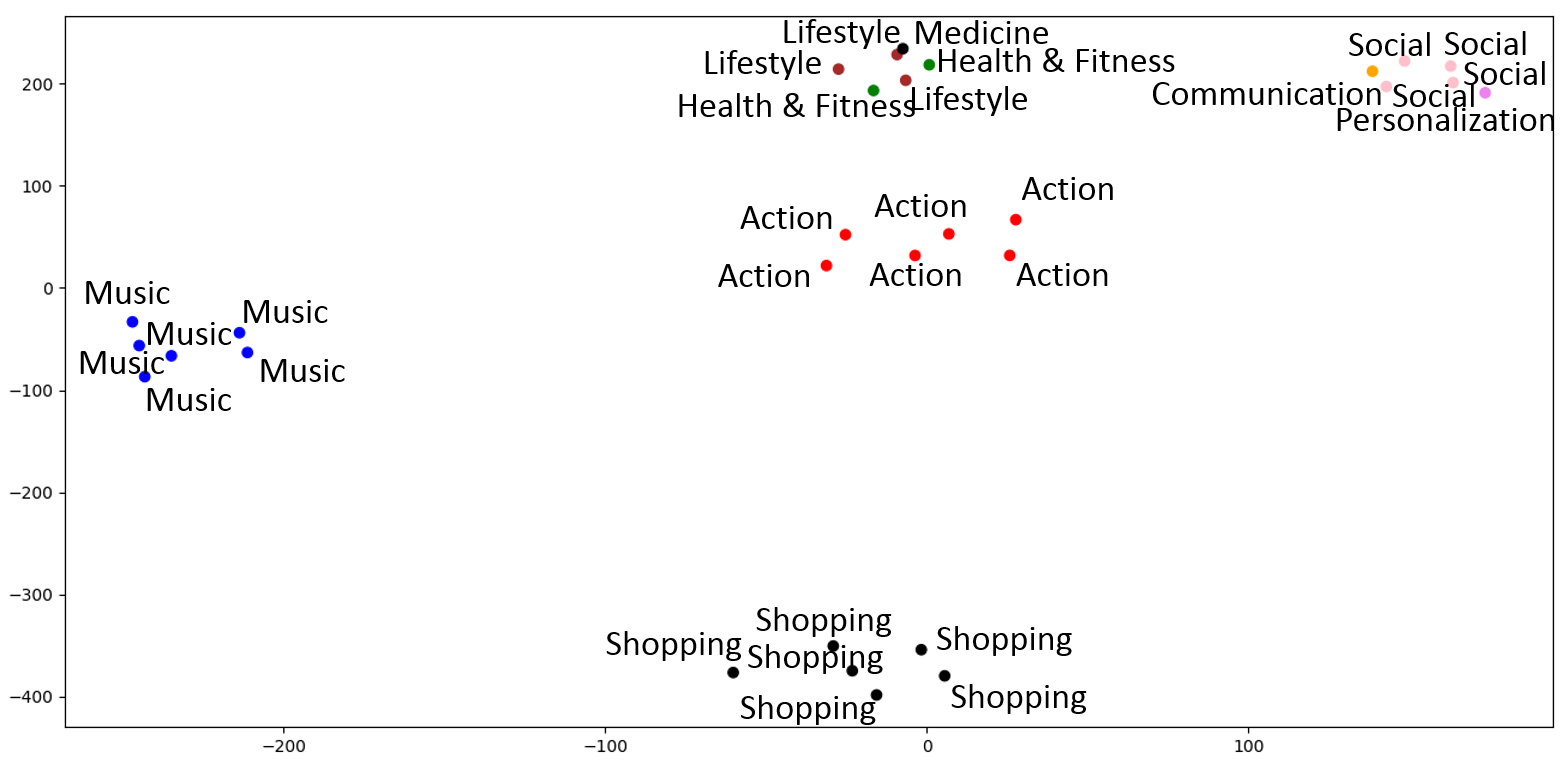}
	\caption{t-SNE projection for applications and their categories given in Table~\ref{table:NNA}}
	\label{fig:tsne}
\end{figure}

Not only quantitatively, but also qualitatively, the proposed application embeddings have proved its worth. It is not always the case that user provides a query term for app retrieval, a user may want to get some recommendations based on a particular app. Similar to \cite{biswas2017mrnet}, Nearest neighbor analysis acts as recommender tool to get the closest match from a specific application. This recommendation system being retrieval based, $RelEmb$ can be employed as application embedding. Qualitative results for nearest neighbor analysis using $RelEmb$ (equation ~\ref{eq:relemb}) are shown in Table~\ref{table:NNA}. From the results it can be seen that for application of a particular category (query), the closest matched apps mostly belong to the same category. For these selected applications, $RelEmb$ embeddings are plotted in a two-dimensional space using t-SNE visualization \cite{maaten2008visualizing} (shown in Figure~\ref{fig:tsne}). The visualization also shows accurate grouping for different categories of applications. This section shows the effectiveness of application embedding for extracting similar apps.
\begin{table*}
	\caption{Qualitative results with Nearest Neighbors Analysis using $RelEmb$ }
	\label{table:NNA}
	\centering
	
	\begin{tabular}{>{\centering\arraybackslash}p{20mm}>{\centering\arraybackslash}p{18mm}>{\centering\arraybackslash}p{18mm}>{\centering\arraybackslash}p{18mm}>{\centering\arraybackslash}p{18mm}>{\centering\arraybackslash}p{18mm}}
		\toprule
		\textbf{Application} & \multicolumn{5}{c}{\textbf{Predicted Nearest Applications - Name (Category)}}\\
		\cmidrule{2-6}
		cops robbers jail break \textbf{(Action)} & survival hungry games (Action) & dead target zombie (Action) & wanted survival games (Action) & cube duty ghost blocks (Action) & orange block prison break (Action)\\
		\cmidrule{1-6}
		
		real piano \textbf{(Music)}
		&drum (Music)
		&real guitar (Music)
		&congas bongos (Music)
		& tabla (Music)
		& hip hop beatz (Music)\\
		\cmidrule{1-6}
		
		relax rain nature sounds \textbf{(Lifestyle)}
		&relax night nature sounds (Lifestyle)
		&relax forest nature sounds (Lifestyle)
		&white noise (Health \& Fitness)
		&pain depression (Medical)
		& melodies sleep yoga (Health \& Fitness)
		\\
		\cmidrule{1-6}
		love phrases images \textbf{(Social)}
		&top good night images (Social)
		&top good morning images (Social)
		&themes classic (Personalization)
		&status messages (Social)
		&video chat friendcaller (Communication)
		\\
		\cmidrule{1-6}
		kroger \textbf{(Shopping)}
		&ralphs (Shopping)
		&smith (Shopping)
		&king soopers (Shopping)
		&fred meyer (Shopping)
		&dillons (Shopping)\\    
		\bottomrule
	\end{tabular}
\end{table*}
\subsection{Application Categorization}
\label{categorization}

Under Application categorization, classification and clustering are the two subdomains, which are most useful. Both classification and clustering use application embeddings as the feature vector for training. Classification is a supervised learning algorithm which uses the document (or application) embedding as feature vector and the application category as the output variable. Trained application classifier can act as black box to predict the category of any new application. On the other hand, clustering is an unsupervised approach based on just the feature vectors. It can be used for grouping similar applications together, which has many practical use-cases like folder creation for grouping similar apps in mobile. 
\subsubsection{App Classification}
\label{classification}
\begin{sloppypar}
Decision tree and multi-class SVM are used to train the classifiers with features being application embedding and labels as application category. The considered TIMAN dataset has application data belonging to 41 categories. To evaluate the performance of the proposed $RelEmb$ and $RelEmbAE$, the dataset is divided into training and test splits using K-Fold cross validation (k = 5) Multi-class classification using Doc2Vec embedding approach \cite{le2014distributed} is used as baseline for comparison. Evaluation results are presented in Table~\ref{table:Classification} and the reported F1-score is computed by averaging F1-score from all 5 groups (after K-Fold). For multi-class SVM classifier, the F1-score (in percentage) for $RelEmbAE$ is 43.866 outperforming the baseline Doc2Vec, which gives the score of 32.55. Similarly, for Decision tree the F1-score (in percentage) for $RelEmbAE$ is 21.332 whereas for baseline Doc2Vec it is 8.536. The results indicate the better performance of $RelEmbAE$ over Doc2Vec and also $RelEmb$. 
\end{sloppypar}
\begin{table*}
	\caption{Evaluation results on App Classification (41 categories)}
	\label{table:Classification}
	\centering
	
	\begin{tabular}{>{\centering\arraybackslash}p{40mm}>{\centering\arraybackslash}p{18mm}>{\centering\arraybackslash}p{18mm}>{\centering\arraybackslash}p{18mm}}
		\toprule
		& \multicolumn{3}{c}{\textbf{F1 Score (Percentage)}}\\
		\cmidrule{2-4}
		\textbf{Classifier} & Doc2Vec & RelEmb & RelEmbAE\\
		\midrule
		Decision Tree     &8.536 &10.351 &\textbf{21.332}\\
		Multi-class SVM &32.550 &40.557 &\textbf{43.866}\\
		\bottomrule
	\end{tabular}
	%
\end{table*}
\subsubsection{App Clustering}
\label{clustering}
K-mean and DBSCAN clustering algorithms are used to evaluate the performance of application embedding for the task of clustering. The results with proposed $RelEmb$ and $RelEmbAE$ are compared with the baseline Doc2Vec embeddings. Two different metrics, such as Silhouette \cite{rousseeuw1987silhouettes} and Davies Bouldin \cite{davies1979cluster} scores are used to compare the results and the evaluation results on App Clustering are shown in Table~\ref{table:Clustering2}. It is known that the value of silhouette score should be more closer to 1 for an accurate clustering. The scores for $RelEmbAE$ are positive indicating better clustering in comparison to Doc2Vec for which the scores are negative, indicating wrongly assigned clusters. For Davies Bouldin score, a value closer to zero means a better separation between the clusters indicating superior performance of clustering. The numbers for both the clustering techniques indicate that $RelEmbAE$ outperforms existing Doc2Vec embedding by significant margins.

As discussed in section~\ref{docemb}, $RelEmbAE$ is more suitable for categorization tasks whereas $RelEmb$ performs well for retrieved-based tasks such as query expansion. For classification and clustering tasks, $RelEmb$ still beats the state-of-the-art Doc2Vec approach showcasing the significance of initial relevance-based word representations learned in section~\ref{wordrepr} and denoted by $WordRepr$. 
\begin{table*}
	\caption{Evaluation results on Clustering Techniques}
	\label{table:Clustering2}
	\centering	
	\begin{tabular}{>{\centering\arraybackslash}p{22mm}>{\centering\arraybackslash}p{20mm}>{\centering\arraybackslash}p{20mm}>{\centering\arraybackslash}p{20mm}>{\centering\arraybackslash}p{20mm}}
		\toprule
		& \multicolumn{2}{c}{\textbf{DBSCAN}} & \multicolumn{2}{c}{\textbf{k-Means (k = 41)}}\\
		\cmidrule{2-5}
		\textbf{Embeddings} & Silhouette Score & Davies Bouldin Score & Silhouette Score & Davies Bouldin Score\\
		\midrule
		Doc2Vec     &-0.311 &2.895 &-0.0903 &4.84 \\
		RelEmb     &0.371 &2.275 &-0.061 &4.376 \\
		RelEmbAE    &\textbf{0.939} &\textbf{1.98} &\textbf{0.289} &\textbf{0.8947}\\
		\bottomrule
	\end{tabular}
\end{table*}

\section{Conclusions and Future Work}
\label{conclusion}
 In this paper, word embeddings are learned with a neural network architecture using the description of mobile applications. These embeddings are developed by keeping in mind the increasing usage of mobile applications and the difficulties faced to find out relevant ones from a large collection. The learning of word embeddings is carried out based on the notion of relevance. The results show that the learned word embeddings are effective for query expansion task eventually making the search experience more user friendly. In addition, the learned word embeddings are also aggregated to find RelEmbAE and RelEmb- distributed and dense representations of mobile application. These embeddings have outperformed Doc2vec on tasks like app classification and clustering. In future, other parameters like application reviews, ratings, etc. can be used to make the application embedding much more rich and descriptive. Another extension of this work can be to analyze the use of application embeddings in the field of query intent detection, query classification which are current topics of research in Information Retrieval. Although the work in this paper is focused on tasks related to mobile applications, the same techniques can be applied to any generic scenarios of Information Retrieval.

\bibliographystyle{splncs}
\bibliography{mainPaper}
\end{document}